# Photo-physical properties of He-related color centers in diamond

G. Prestopino[1], Marco Marinelli[1], E. Milani[1], C. Verona[1], G. Verona-Rinati[1]
P. Traina[2], E. Moreva[2], I.P. Degiovanni[2], M. Genovese[2]
S. Ditalia Tchernij[3,4], F. Picollo[3,4], P. Olivero[3,4] J. Forneris[4,3*]

[1] INFN–Dipartimento di Ingegneria Industriale, Università di Roma "Tor Vergata," Via del Politecnico 1, Roma 00133, Italy
[2] Istituto Nazionale di Ricerca Metrologica (INRiM), Strada delle Cacce 91, 10135 Torino, Italy
[3] Physics Department and "NIS" Inter-departmental Centre - University of Torino, via P. Giuria 1, 10125 Torino, Italy
[4] Istituto Nazionale di Fisica Nucleare (INFN) Sez. Torino, Via P. Giuria 1, 10125 Torino, Italy

* corresponding author. email: forneris@to.infn.it

## Abstract

Diamond is a promising platform for the development of technological applications in quantum optics and photonics. The quest for color centers with optimal photo-physical properties has led in recent years to the search for novel impurity-related defects in this material. Here, we report on a systematic investigation of the photo-physical properties of two He-related (HR) emission lines at 535 nm and 560 nm created in three different diamond substrates upon implantation with 1.3 MeV He$^+$ ions and subsequent annealing. The spectral features of the HR centers were studied in an "optical grade" diamond substrate as a function of several physical parameters, namely the measurement temperature, the excitation wavelength and the intensity of external electric fields. The emission lifetimes of the 535 nm and 560 nm lines were also measured by means of time-gated photoluminescence measurements, yielding characteristic decay times of (29 ± 5) ns and (106 ± 10) ns, respectively. The Stark shifting of the HR centers under the application of an external electrical field was observed in a CVD diamond film equipped with buried graphitic electrodes, suggesting a lack of inversion symmetry in the defects' structure. Furthermore, the photoluminescence mapping under 405 nm excitation of a "detector grade" diamond sample implanted at a 1×10$^{10}$ cm$^{-2}$ He$^+$ ion fluence enabled to identify the spectral features of both the HR emission lines from the same localized optical spots. The reported results provide a first insight towards the understanding of the structure of He-related defects in diamond and their possible utilization in practical applications.

## MAIN TEXT

Color centers in diamond are appealing physical systems for application in emerging quantum technologies. In recent years, a growing interest in this research field led to the discovery, investigation and fabrication of several classes of impurity-related defects [1-8].

The formation of optically active defects in diamond upon He implantation followed by annealing in vacuum was recently reported [9]. The spectral features of these defects consist of two narrow emission lines at 536 nm and 560.5 nm, and a phonon sideband in the 572-630 nm spectral range. Following their initial observation in cathodoluminescence (CL) [10, 11], their optical activity was investigated in both photoluminescence (PL) and electroluminescence (EL) [9, 12].

The need of understanding the possible formation of stable optically active centers, whose current (and still tentative) attribution is based on He incorporation in the diamond lattice, motivates a further investigation of their opto-physical properties. More specifically, the centers have so far been attributed to the He-vacancy complex [13], and density functional theory calculations demonstrated that both this structure and the interstitial He defect could result in stable complexes in the diamond lattice [14]. Apart from this limited body of works, a systematic set of experimental results on the characterization of the opto-physical properties of He-Related (HR, from now on) centers is still missing.

In this letter, we present the results of an extensive characterization of single crystal diamond samples incorporating HR centers, with the purpose of gaining a deeper insight in their opto-physical properties. Their PL spectral features were investigated as a function of temperature, excitation wavelength and intensity of external electrical fields. Additionally, the single-photon-sensitive PL confocal microscopy of a diamond substrate containing HR centers at very low densities enabled their preliminary mapping, thus providing indications on the spatial distribution/correlation of HR emission lines from the same complexes.

The reported measurements were performed on a set of three He-implanted samples: a type-IIa synthetic single-crystal "optical grade" diamond plate ($3\times3\times0.3$ mm$^3$ in size) by ElementSix (Sample #1); a single-crystal diamond film grown at the laboratories of "Tor Vergata" University in Rome on a commercial high-pressure high-temperature (HPHT) Ib diamond substrate (Sample #2); a type-IIa "detector grade" diamond plate by IIA Technologies (Sample #3) with a nominal substitutional N concentration of <1 ppb. $500\times500$ μm$^2$ regions were irradiated with a 1.3 MeV He$^+$ beam at $2\times10^{16}$ cm$^{-2}$ and $1\times10^{10}$ cm$^{-2}$ fluences on Sample #1 and #3, respectively. Sample #2 was fabricated with graphitic micro-electrodes by means of deep ion beam lithography technique [15]. Fabrication details are discussed in Refs. [12, 16]. To summarize, pairs of electrodes with ~12 μm spacing (see Figure 3(a)) were fabricated through a 1.8 MeV He$^+$ microbeam ($1.5\times10^{17}$ cm$^{-2}$ implantation fluence, i.e. above the graphitization threshold [15]). The utilization of a finite-size ion microbeam for the microchannels fabrication resulted in the creation of a "halo" of HR defects surrounding the electrodes, with an estimated implantation fluence of ~$10^{15}$ cm$^{-2}$ at the center of the inter-electrode gap. All substrates were subsequently annealed at 1000 °C for 2 hours in vacuum to induce the formation of HR defects.

PL measurements as a function of temperature and excitation wavelength were performed on Sample #1. The same experimental system described in Ref. [17] was used. Briefly, Sample #1 was placed on a cold finger inside a double-stage cryostat, allowing to vary its temperature in the 25-300 K range. A 3D movable micrometric stage enabled the positional control of the optical excitation (spot beam area $4.5\times10^{-4}$ cm$^2$), which was in turn provided by a 5 ns pulsed laser (20 Hz repetition rate) with a tunable wavelength in the 210-500 nm range. PL spectra were acquired by a spectrograph/gated multichannel-plate CCD camera system, enabling acquisitions with 2 ns time resolution (600 lines mm$^{-1}$ and 1200 lines mm$^{-1}$ grating, respectively in what will be referred as "low" and "high spectral resolution" modes in the following). Typical low-resolution PL spectra acquired with a 440 nm excitation wavelength at 25 K, 100 K, 200 K, and 300 K are displayed in **Figures 1(a-d)**, respectively. The most prominent features consist of the two HR zero-phonon lines (ZPL) at 535.2 nm (HR1) and 559.7 nm (HR2). The ZPL emission of the neutral nitrogen-vacancy center (NV$^0$) at 575 nm is also displayed to provide a comparison between the HR emission properties and the most studied optical center in diamond. The relative intensities of the emission lines cannot be taken as a direct indication of their respective quantum efficiencies, since the PL detection occurred with a fixed 70 ns delay with respect to the laser excitation, thus affecting the collected signals characterized by different emission lifetimes.

An additional emission doublet (not visible at room-temperature, due to a higher PL background) consisted of two peaks centered at 562.7 nm and 563.3 nm. This emission was previously reported both in PL and EL regimes [18-19] in samples irradiated with MeV C ions, and cannot be therefore related to He-containing centers. It was tentatively attributed to interstitial defects according to previous works [11, 20].

High-resolution spectra of the HR1 and HR2 ZPLs, acquired at 100 K and normalized to the number of excitation photons per laser pulse (**Figures 1(e)** and **1(f)**, black lines) did not exhibit any fine structure, within the ∼0.06 nm spectral resolution of the setup. Additional measurements carried out with a 220 nm excitation wavelength (**Figures 1(e)** and **1(f)**, red lines) demonstrate that the aforementioned spectral features are active under excitation at shorter wavelengths. The higher emission intensity observed under 220 nm excitation is a consequence of the diamond band-to-band photon absorption, as shown in details in Fig. 2e.

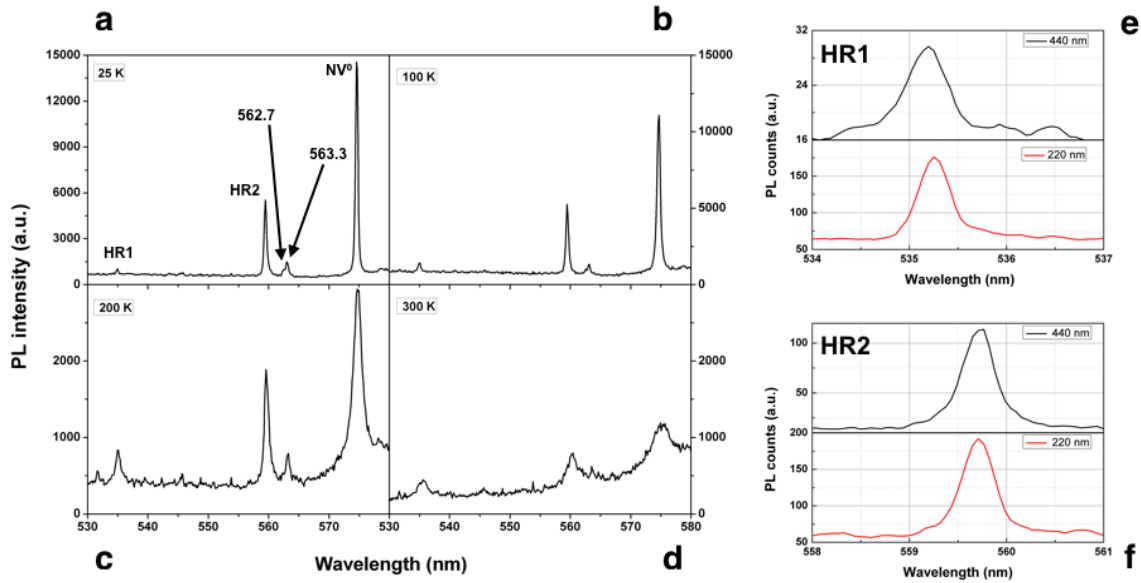

**Figure 1:** PL spectra at different temperatures under 440 nm excitation wavelength: (**a**) 25 K; (**b**) 100 K; (**c**) 200 K; (**d**) 300 K. High resolution spectra of (**d**) HR1 and (**e**) HR2 emission lines measured at 100 K under 440 nm (black lines) and 220 nm (red lines) excitation wavelengths.

The dependence of the PL emission intensities from temperature is reported in **Figure 2(a)**, in which the integrated peaks intensities after background subtraction are normalized to the respective values measured at 25 K. The intensity decrease in the HR2 and $NV^0$ emissions at increasing temperature is a commonly observed feature in these systems, which is conveniently described by the higher probability for the radiative emission to involve excited vibrational states. Notably, on the contrary, the HR1 emission intensity increases with the temperature, possibly indicating that such emission is associated with a phonon-assisted process.

**Figure 2(b)** shows the red-shift of HR1 (black filled squares) and HR2 (red filled circles) ZPLs, as a function of the fourth power of the temperature ($T^4$), together with the same data for the $NV^0$ emission (blue filled triangles) for sake of comparison. The linearization of the energy shift dependence from $T^4$ is consistent with the consolidated interpretative model based on the contraction of the diamond lattice and on the electron-phonon coupling with hard phonon modes [11,21-22].

The dependence of the spectral widths (i.e. peak FWHM), in nanometer units, of the HR1 and HR2 emissions from the third power of the temperature ($T^3$) is reported in **Figure 2(c)**. PL data for the $NV^0$ center are reported again for comparison. The $T^3$ dependence of the FWHM peaks, highlighted by the fitting lines, is associated with the softening of the chemical bonds of the defect complex and results in a stronger coupling of its excited state with lattice phonons [23]. The FWHM increase versus temperature is less pronounced for HR centers with respect to the $NV^0$ emission, indicating that the former are characterized by a lower phonon coupling, as confirmed by their weaker phonon sidebands.

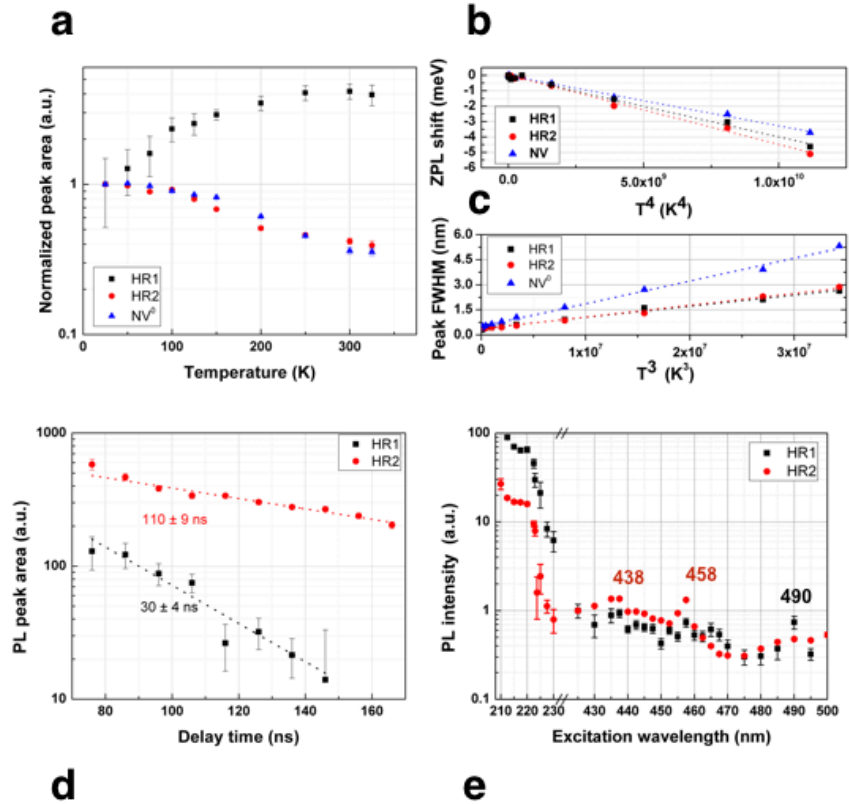

**Figure 2:** (**a**) PL emission intensity of HR1, HR2, and $NV^0$ centers as a function of sample temperature. Each set of data is normalized to the corresponding value measured at 25 K. (**b**) Red-shift of the HR1, HR2 and $NV^0$ ZPLs as a function of $T^4$. (**c**) PL peak widths (FWHM) as a function of $T^3$. The dashed lines indicate the respective linear best fits. **d**) Typical PL intensity decay chronograms of HR1 and HR2 emissions at 100 K under 440 nm laser excitation. **e**) Dependence of the PL emission intensity at 100 K from HR1 and HR2 centers as a function of laser excitation wavelength in the 210-510 nm range.

The PL emission lifetime of the HR1 and HR2 centers was estimated in Sample #1 at 100 K temperature under 440 nm excitation wavelength (3.5 mW average power), by measuring the time-gated PL emission with respect to pulsed excitation. Typical PL intensity decay chronograms of the HR1 and HR2 emissions are reported in **Figure 2(d)**, together with the corresponding exponential fitting functions and the calculated decay times. From the statistical analysis of a set of ~10 independent measurements, the average emission lifetimes were estimated as $\tau_1 = (29 \pm 5)$ ns and $\tau_2 = (106 \pm 10)$ ns for the HR1 and HR2 emissions, respectively. These values were fully compatible with further measurements obtained under 220 nm excitation (not shown here) and did not display significant variations in the 25-100 K temperature range.

The dependence of HR1 and HR2 PL intensity on laser wavelength (210 - 510 nm range, 100 K temperature) is shown in **Figure 2(e)**. The measured PL intensities were normalized to the wavelength-dependent photons number per each laser excitation pulse. The FWHM of the peaks, as well as their respective center positions, did not display any significant variations over the considered spectral ranges. As an overall trend, HR1 and HR2 centers display decreasing PL emission intensities at increasing excitation wavelengths. An intensity maximum at 490 nm was apparent for the HR1 emission. Conversely, the HR2 line exhibited two maxima at 438 nm and 458 nm. No significant mutual correlations between the HR1 and HR2 PL intensity variations were observed in the afore-mentioned range, therefore no conclusions on the nature of the two emissions (e.g., possible associations with different charge states of the same complex) could be inferred. Furthermore, no apparent center ionization, associated with a sudden decrease of the PL intensity at high energies was observed down to 405 nm excitation wavelength.

Remarkably, the data acquired under UV excitation revealed a sudden increase in the PL emission at wavelengths lower than 225 nm (corresponding to the band-edge absorption related to the 5.5 eV band gap). This indicates that the recombination of electron-hole pairs generated upon band-to-band photon absorption in the UV provides a further excitation channel for the HR centers emission, and thus confirms the role of the defects as electrically active traps in the material.

The Stark shifting of the HR centers under the application of an external electrical field was investigated on Sample #2. PL spectra under different applied electrical bias voltages were acquired with the same setup described in [19], consisting of a Horiba Jobin Yvon HR800 Raman micro-spectrometer with a 1800 groves mm$^{-1}$ diffraction grating and a 532 nm laser excitation (21.6 mW power on the sample surface). The Stark shift induced by static electrical fields to the HR1 and HR2 transitions was investigated at room temperature by applying a bias voltage difference in the 0-500 V range to a small (i.e. ~10×15 μm$^2$) region comprised between sub-superficial graphitic electrodes [12]. The position of the laser excitation spot (~3 μm in size) is highlighted by the red circle in **Figure 3(a).** It is worth remarking that the EL emission observed in previous works in the same regions [12] was quenched by the intense laser excitation, which prevented the device from reaching the trap-filling-induced transition to the Poole-Frenkel conduction regime [18-19].

The variations of the HR1 and HR2 PL emission spectra at increasing voltages are displayed in **Figures 3(b)** and **3(c)**, respectively. Both peaks exhibit a red-shift as a consequence of the applied electric field. In **Figure 3(d)**, the spectral shift is reported against the bias voltage for both the HR ZPLs. From a linear fitting, the Stark shift effect was quantified with slope values of -(1.9±0.2)×10$^{-9}$ eV cm V$^{-1}$ and -(1.8 ± 0.2)×10$^9$ eV cm V$^{-1}$ for the HR1 and HR2 lines, respectively.

The observation of a Stark shift of the HR1 and HR2 centers indicates a lack of inversion symmetry in the defects structure [11]. This fact is in agreement with the interpretation of the emission as associated to He-V defects, rather than to interstitial He, as proposed in Refs [11, 13-14].

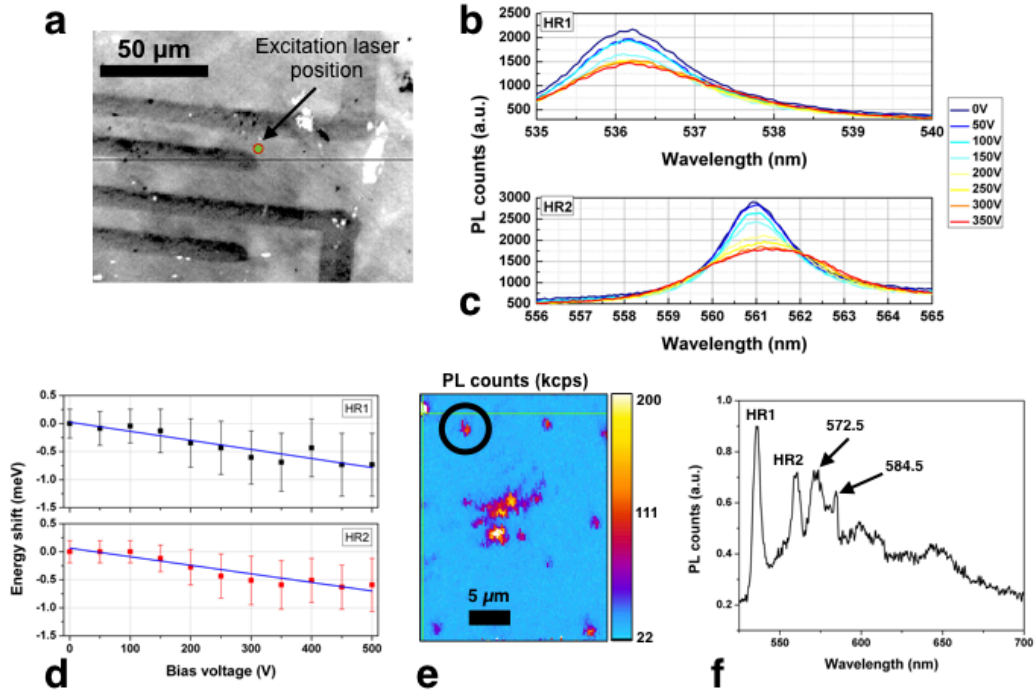

**Figure 3: a)** Optical micrograph of the graphitic micro-electrodes in Sample #2. The red circle indicates the position of the laser excitation spot. PL spectra **b)** of the HR1 center, and **c)** of the HR2 center under increasing applied bias voltage. **d)** Red shift in the HR1 and HR2 emission at increasing bias. The blue lines indicate the respective linear best fits. **e)** Typical PL map acquired from Sample #3 under 405 nm CW laser excitation. **f)** PL spectrum acquired from the spot circled in black in Figure (e).

Confocal PL mapping of HR defects was performed on Sample #3 under 405 nm CW laser excitation (15 mW power at the sample surface) using a single-photon-sensitive setup [9] with photon collection limited by a set of filters to the 500-570 nm spectral range. PL spectra were acquired through a single-grating monochromator (1600 grooves mm$^{-1}$, 600 nm blaze). The sample, containing a low density of both irradiation-induced HR and native NV centers, was investigated to gain an insight on the spatial distribution and mutual correlations of HR1 and HR2 emissions. A typical confocal PL map is reported in **Figure 3(e)**. It exhibits several bright isolated spots with a size compatible with that of isolated emitters embedded among larger luminescent regions, while the substrate is characterized by a low fluorescence background. Remarkably, all of the above-mentioned regions (i.e. both the isolated spots and the larger agglomerates) have similar spectral features. A typical PL spectrum from one of these spots (circled in black in **Figure 3(e)**) is shown in **Figure 3(f)**.

Besides the HR1 and HR2 lines, the spectrum displays additional spectral features, such as two PL peaks at 572.5 nm and 584.5 nm, together with the HR2 phonon replica in the 580-630 nm range [9]. The present measurements performed on a substrate with a low concentration of NV centers suggest that the origin of such spectral features is related to the HR emissions. Despite the fact that preliminary measurements of the second-order autocorrelation function [24] did not enable to identify single-photon emission, our point measurements indicate that both HR1 and HR2 emissions (and relevant lines) come from the same spot, i.e. possibly from the same defect complex [10]. On the other hand, it is worth noting that the PL emission from the isolated spots did not exhibit any intensity variations over the ~500 s measurement duration. Therefore, at least at these spatial (i.e. ~1 μm) and temporal (~10$^{-1}$ s) resolutions, the attribution of the HR1 and HR2 emissions to different charge states of the same defect could not be directly verified.

We presented a detailed characterization of the photo-physical properties of HR centers ensembles fabricated in diamond upon MeV He implantation. We showed that the centers did not exhibit any fine structure at temperatures as low as 25 K, and we demonstrated that the relevant emission features (ZPL shift, peaks FWHM) follow the typical trend against the temperature of those typical of stable defects in the diamond lattice. Contrarily to the HR2 centers, we observed a PL increase of the HR1 emission intensity at increasing temperatures, which could be indicative of the presence a phonon-assisted radiative process. The emission lifetimes at 100 K were quantified as (29 ± 5) ns and (106 ± 10) ns for the HR1 and HR2 centers, respectively.

The HR centers could be effectively excited under both visible and UV radiation, and exhibited intense luminescence at energies higher than the diamond band gap, thus confirming their activity in CL and EL regime. The HR centers exhibited a Stark shift effect under applied electrical bias, thus ruling out the inversion symmetry from their defect structure. Finally, PL confocal mapping under 405 nm excitation revealed that the HR1 and HR2 emissions arise from the very same luminescent spots in the sample. Future studies will focus on the investigation of the HR emission at the single-photon emitter level, in order to clarify the possible attribution of the HR1 and HR2 lines to different charge states of the same defect complex. In principle, the HR1 emission lifetime should enable the detection of luminescence from individual defects, while the HR2 center is almost three times weaker in emission due to its relatively long lifetime. Overall, the narrow ZPL emission of HR centers with small phonon coupling and the photo-excitability under a wide spectral range, combined with the large availability of focused He beams for fabrication [16, 25-27] represent promising aspects for the application of HR defects in quantum technologies.

## Acknowledgements

This research activity was supported by the following projects: "DIESIS " project funded by the Italian National Institute of Nuclear Physics (INFN) - CSN5 within the "Young research grant " scheme; Coordinated Research Project F11020 funded by the International Atomic Energy Agency (IAEA); Progetto Premiale 2014 "Q-SecGroundSpace" funded by MIUR. MeV He implantations were performed within the "Dia.Fab. " experiment at the INFN-LNL laboratories.

# References

[1] J.O. Orwa, A.D. Greentree, I. Aharonovich, A.D.C. Alves, J. Van Donkelaar, A. Stacey, and S. Prawer, J. Lumin. **130**, 1646 (2010).

[2] S. Pezzagna, D. Rogalla, D. Wildanger, J. Meijer, and A. Zaitsev, New J. Phys. **13**, 035024 (2011).

[3] I. Aharonovich, S. Castelletto, D.A. Simpson, C.-H. Su, A.D. Greentree, and S. Prawer, Rep. Prog. Phys. **74**, 076501 (2011).

[4] T.M. Babinec, B.J.M. Hausmann, M. Khan, Y. Zhang, J.R. Maze, P.R. Hemmer, and M. Lončar, Nat. Nanotechnol. **5**, 195 (2010).

[5] M. Leifgen, T. Schröder, F. Gädeke, R. Riemann, V. Métillon, E. Neu, C. Hepp, C. Arend, C. Becher, and K. Lauritsen, New J. Phys. **16**, 023021 (2014).

[6] T. Iwasaki, F. Ishibashi, Y. Miyamoto, Y. Doi, S. Kobayashi, T. Miyazaki, K. Tahara, K.D. Jahnke, L.J. Rogers, B. Naydenov, F. Jelezko, S. Yamasaki, S. Nagamachi, T. Inubushi, N. Mizuochi, and M. Hatano, Sci. Rep. **5**, 12882 (2015).

[7] D. Gatto Monticone, P. Traina, E. Moreva, J. Forneris, P. Olivero, I.P. Degiovanni, F. Taccetti, L. Giuntini, G. Brida, G. Amato, and M. Genovese, New J. Phys. **16**, 053005 (2014).

[8] R. John, J. Lehnert, M. Mensing, D. Spemann, S. Pezzagna, and J. Meijer, New J. Phys. **19**, 053008 (2017).

[9] J. Forneris, A. Tengattini, S. Ditalia Tchernij F. Picollo, A. Battiato, P. Traina, I.P. Degiovanni, E. Moreva, G. Brida, V. Grilj, N. Skukan, M. Jakšić, M. Genovese, and P. Olivero, J. Lumin. **179**, 59 (2016).

[10] V.D. Tkachev, A.M. Zaitsev, and V.V. Tkachev, Phys. Stat. Sol. (b) **129**, 129 (1985).

[11] A.M. Zaitsev, Optical properties of Diamond (Springer, New York, 2001).

[12] J. Forneris, A. Battiato, D. Gatto Monticone, F. Picollo, G. Amato, L. Boarino, G. Brida, I.P. Degiovanni, E. Enrico, M. Genovese, E. Moreva, P. Traina, C. Verona, G. Verona Rinati, and P. Olivero, Nucl. Instr. Meth. B **348**, 187 (2015).

[13] B. Dischler , Handbook of Spectral Lines in Diamond Vol.1: Tables and Interpretations (Springer, Berlin, 2012).

[14] J.P. Goss, R.J. Eyre, and P.R. Briddon, Phys. Rev. B **80**, 085204 (2009).

[15] F. Picollo, A. Battiato, E. Bernardi, L. Boarino, E. Enrico, J. Forneris, D. Gatto Monticone, and P. Olivero, Nucl. Instr. Meth. B **348**, 199 (2015).

[16] J. Forneris, V. Grilj, M. Jakšić, A. Lo Giudice, P. Olivero, F. Picollo, N. Skukan, C. Verona, G. Verona-Rinati, and E. Vittone, Nucl. Instr. Meth. B **306**, 181 (2013).

[17] M.G. Donato, G. Messina, G. Verona Rinati, S. Almaviva, G. Faggio, M. Marinelli, E. Milani, G. Prestopino, S. Santangelo, P. Tripodi, and C. Verona, J. Appl. Phys. **106**, 053528 (2009).

[18] J. Forneris, P. Traina, D. Gatto Monticone, G. Amato, L. Boarino, G. Brida, I.P. Degiovanni, E. Enrico, E. Moreva, V. Grilj, N. Skukan, M. Jakšić, M. Genovese, and P. Olivero, Sci. Rep. **5**, 15901 (2015).

[19] J. Forneris, S. Ditalia Tchernij, A. Tengattini, E. Enrico, V. Grilj, N. Skukan, G. Amato, L. Boarino, M. Jakšić, and P. Olivero, Carbon **113,** 76 (2017).

[20] J.W. Steeds, S. Charles, T.J. Davis, A. Gilmore, J. Hayes, D. Pickard, and J.E. Butler, Diam. Relat. Mater. **8**, 94 (1999).

[21] K. Dragounova, Z. Potucek, S. Potocky, Z. Bryknar, and A. Kromka, J. Elec. Eng. **68**, 74 (2017).

[22] M.W. Doherty, V.M. Acosta, A. Jarmola, M.S.J. Barson, N.B. Manson, D. Budker, and L.C.L. Hollenberg, Phys. Rev. B **90**, 041201 (2014).

[23] V. Hizhnyakov, H. Kaasik, and I. Sildos, Phys. Stat. Sol. (b) **234**, 644 (2002).

[24] A. Beveratos, S. Kühn, R. Brouri, T. Gacoin, J.-P. Poizat, and P. Grangier, Eur. Phys. J. D **18**, 191 (2002).

[25] A.H. Piracha, K. Ganesan, D.W.M. Lau, A. Stacey, L.P. McGuinness, S. Tomljenovic-Hanic, and S. Prawer, Nanoscale **8**, 6860 (2016).

[26] Z. Huang, , W.-D. Li, C. Santori, V.M. Acosta, A. Faraon, T. Ishikawa, W. Wu, D. Winston, R.S. Williams, and R.G. Beausoleil, Appl. Phys. Lett. **103**, 081906 (2013).

[27] D. McCloskey, D. Fox, N. O'Hara, V. Usov, D. Scanlan, N. McEvoy, G.S. Duesberg, G.L.W. Cross, H.Z. Zhang, and J.F. Donegan, Appl. Phys. Lett. **104**, 031109 (2014).